\documentclass[12pt]{iopart}

\expandafter\let\csname equation*\endcsname\relax
\expandafter\let\csname endequation*\endcsname\relax 
\usepackage{amsmath}  
\usepackage{amsfonts} 
\usepackage{amssymb}
\usepackage{physics} 
\usepackage{iopams}  
\usepackage{graphicx}
\usepackage[breaklinks=true,colorlinks=true,linkcolor=blue,urlcolor=blue,citecolor=blue]{hyperref}
\usepackage[utf8]{inputenc}
\usepackage[caption = false]{subfig}
\usepackage{float}
\usepackage{amsmath,amsfonts,amsthm,bm}
\usepackage{braket}
\usepackage{subfig}

\begin{document}
\title{Bound States of Central Force System in Special Relativity}

\author{Iwan Setiawan$^{1}$, Ryan Sugihakim$^{2}$, and Bobby Eka Gunara$^2$}
\address{$^1$ Department of Physics Education, University of Bengkulu, Kandang Limun, Bengkulu 38371, Indonesia}
\address{$^2$ Department of Physics, Institut Teknologi Bandung, Jalan Ganesha 10, Bandung 40132, Indonesia}
\ead{iwansetiawan@unib.ac.id}
\date{\today}

\begin{abstract}
In this paper we consider the central force problem in the special theory of relativity. We derive the special relativistic version of the Binet equation describing the orbit of a body. Then, the motion of a planet in a solar-like system where the gravitational potential is modified by adding a term that is  proportional to the inverse of the square of the radial coordinate is discussed. Using perturbative method,  we obtain the explicit orbital function. At the end, the orbital precession of the planet and  its relation to the result from general relativity are also discussed.
\end{abstract}

\section{INTRODUCTION}

One of the common situations in physics is the motion of a body under influence of the central force in which the magnitude of the force depends only on the radial coordinate.  The problem is relevant to realistic conditions such as the motion under either gravitational force or electrostatic force. Many standard textbooks in classical mechanics discuss the central force problem such as two-body system to solve the motion of interacting objects, see for example, \cite{goldstein_classical_2001, arya_introduction_1990}. We can say that this problem is essential to physics students improving their competence.

In the non-relativistic mechanics, the central force problem can be reduced to second order ordinary differential equation called Binet equation whose solutions are called orbit in which  the position can be viewed as a function of angle in the polar coordinate. Several papers have been published to study this equation. For example, in \cite{Poole2005OrbitsOC, Adkins2007OrbitalPD} some authors considered a macroscopic case where the potential energy has polynomial form which covers the standard Newton gravitational potential energy that is proportional to the inverse of the radial coordinate. They find that the orbit is an implicit function related to a hypergeometric function. At the microscopic level, the authors in \cite{Deeney2014SommerfeldsEA} have studied the Sommerfeld atomic model of Hydrogen-like model. However, we know that this is unrealistic since according to the astronomical observation the relativistic effect cannot be ignored especially in  the motion of a planet near a massive object. Therefore, the Binet equation has to be extended to the relativistic domain.

In the extension to the relativistic mechanics, although  general relativity is the most favourable at the level of macroscopic system, it is still of interest to consider special relativity since it is much simpler than the general theory. In other words, we can study the dynamics of a particle without curving the spacetime. There are several attempt in this direction. For example, in \cite{Nishikawa1990ARE} the author extend the concepts  the central and conservative forces to special relativity. The other authors in \cite{Kumar2011PossiblePR} have tried to extend Bertrand’s theorem to the central force problem in special relativity. They also find that there is a special form of potentials  which admits a stable circular orbit. There are also other authors that have considered this problem at the microscopic level to produce Sommerfeld-Dirac fine structure formula \cite{Granovskii2004SommerfeldFA, Terzis2008ASR}.

According to the above developments, we also attempt  to consider the central force problem in special relativity. First, we review the derivation of  the relativistic Binet equation describing the orbit of a body \cite{Kumar2011PossiblePR, Lemmon2010}. We shortly discuss the stability criterion of the circular orbit. Then, a simple case is considered where the potential is proportional to the inverse of  the radial coordinate. We obtain an explicit orbital function in which the orbital precession of a body appears naturally for elliptical orbits which can be regraded as the relativistic effect \cite{Lemmon2010}, see also the discussion in \cite{Discussstack}. This precession vanishes if the orbit is circular.

Next, at the microscopic level the above results are applied to the case of the hydrogen-like atom. Using the Wilson-Sommerfeld quantization rule, we obtain the energy level of the system so called Sommerfeld-Dirac fine structure formula \cite{Granovskii2004SommerfeldFA, Terzis2008ASR} which can also be derived from Dirac equation in relativistic quantum mechanics \cite{greiner_relativistic_2000, Lu1970DerivationOS}.

At the final part, we discuss a planetary motion in which the potential energy is fine-tuned by adding the term which is proportional to the cubic of of the radial coordinate with arbitrary pre-coefficient. This potential energy has been appeared in the case of a free falling body in Schwarzschild spacetime, see for example, \cite{landau_classical_1980, cheng2009} and Randall-Sundrum model of gravity \cite{Randall_1999}. Using perturbative method, we get an explicit orbital function that admits the orbital precession of the planet and then, we compare it with the result from general relativity. It is worth mentioning that this explicit orbital function  has never been considered in the literature above. Also, we pointed out the second Kepler's laws which differs to the case of Newtonian mechanics.

In this paper, it is convenient to use Lagrangian formulation as a starting point. We use the relativistic Lagrangian that can be found from textbooks such as by Landau and Lifshitz \cite{landau_classical_1980}. We discuss the motion of a body under the influence of the central force using this formulation in section \ref{sec:relativityCentralForce}. In section \ref{sec:HydrogenAtom}, we derive the energy levels of hydrogen-like atom \cite{Granovskii2004SommerfeldFA, Terzis2008ASR, greiner_relativistic_2000, Lu1970DerivationOS}. Then, in section \ref{sec:OrbitPrecession} we discuss the orbit precession and the fine-tuning of model to explain a planet's orbital precession. We add some appendices to support our arguments in the paper.

\section{Special Relativistic Central Force Problem}
\label{sec:relativityCentralForce}

In this section we discuss our general setup of a central force system in special theory of relativity. Then, our discussion will be focus on an attractive force whose scalar potential is proportional to the inverse of the distance.

Let us first consider a relativistic particle with the rest mass $m_0$ moving with speed $v$ under the influence of  a central force that depends only on the radial coordinate $r$ whose potential energy function is given by $V \equiv V(r)$.  This implies that we could reduce the case to two dimensional problem such that using the polar coordinate $(r,\theta)$ for space with coordinate time $t$. We define
\begin{align}
\label{eq:gammaDefined}& \gamma(v) \equiv \qty(1 - \frac{v^2}{c^2})^{-1/2}  ~ .
\end{align}
The Lagrangian of this system is
\begin{align}
L(r,\theta,\dot r, \dot\theta) = -\frac{m_0 c^2}{\gamma(v)} - V(r) ~ .
\end{align}
The dot over a quantity denotes the derivative with respect to $t$.  After some computation we could show that the constants of motion are the angular momentum $p_\theta$ and the total 'mechanical' energy $E$ (see \ref{app:ConstantOfMotion} for details)
\begin{align}
&p_\theta = \gamma m_0 r^2\dot\theta ~ , \\
&\label{eq:EnergyOfSystem} E = \gamma m_0 c^2 + V(r) ~ ,
\end{align}
whereas the radial momentum
\begin{align}
p_r = \gamma m_0 \dot{r} = -p_\theta u' ~ ,
\end{align}
is not constant where $u \equiv 1/r, u'  \equiv \dv*{u}{\theta}$. We note that
\begin{align}
\label{eq:gammaOtherExprs}& \gamma = \qty[1 + \frac{p_\theta^2}{m_0^2 c^2}\qty(u'^2 + u^2)]^{1/2} = \frac{E - V}{m_0 c^2} ~ .
\end{align}
 From \eqref{eq:EnergyOfSystem} and \eqref{eq:gammaOtherExprs}, we obtain then
\begin{align}
\label{eq:prepareBinetCentral}
m_0^2 c^4 = (E - V)^2 - u'^2 p_\theta^2 c^2 - u^2 p_\theta^2 c^2 ~ ,
\end{align}
such that if we differentiate \eqref{eq:prepareBinetCentral} with respect to $\theta$, it leads to the relativistic Binet equation
\begin{align}
 \dv[2]{u}{\theta} + u + \frac{(E - V)}{p_\theta^2 c^2}\dv{V}{u} = 0 ~ . \label{eq:MainBinet2}
\end{align}

Some comments are in order. If $u$ is a positive constant, then we may have a circular orbit. This is the simplest closed orbit which is possible if
the following condition fulfills
\begin{align}
 u_0 =  \frac{( V(u_0) - E)}{p_\theta^2 c^2} \frac{dV}{du}(u_0) ~ , \label{eq:MainBinet2circular}
\end{align}
where $r_0 = 1/u_0$ is the radius of the circular orbit implying $V(u_0) \neq E$ and $u_0$ is not the extremal point of $V(u)$, namely, $\frac{dV}{du}(u_0) \neq 0$. These features distinguish to the case in Newtonian mechanics as discussed in \cite{goldstein_classical_2001}. Moreover, this orbit is stable under small perturbations if it satisfies \cite{Kumar2011PossiblePR}
\begin{align}
\xi \equiv 1-\frac{dJ}{du}(u_0) > 0  ~ , \label{eq:stableorbitcon}
\end{align}
that should be a rational number where
\begin{align}
J \equiv \frac{( V - E)}{p_\theta^2 c^2} \frac{dV}{du}  ~ .
\end{align}
In the case of attractive potentials such as Newtonian gravitational potential and Coulomb potential, the condition \eqref{eq:stableorbitcon} follows
\begin{align}
K^2/p_\theta^2 < c^2  ~ . \label{eq:stableorbitcon1}
\end{align}

\subsection{Orbital Motion in a Special Class of Central Force}
Now, we particularly discuss  the potential energy function whose form is given by
\begin{align}
V(r) = -\frac{K}{r}, \quad K>0 ~ , \label{eq:potentialinvdis}
\end{align}
satisfying \eqref{eq:stableorbitcon1} which appears only in relativistic cases. The potential \eqref{eq:potentialinvdis} describes the attractive force between two particles which is proportional to the inverse of $r^2$. In the case at hand, we may have a solution with closed orbit. Substituting to the equation \eqref{eq:MainBinet2}, we have
\begin{align}
\label{eq:CentralForceBinet}
\dv[2]{u}{\theta} + \qty(1 - \frac{K^2}{p_\theta^2 c^2})u = \frac{KE}{p_\theta^2 c^2} ~ .
\end{align}
Using the condition $u'(0) = 0$, we have the  solution, see the discussion in \cite{Discussstack},
\begin{align}
\label{eq:solutionBinetCentral}
u(\theta) = \frac{1 + e\cos(\kappa\theta)}{a(1-e^2)} ~ ,
\end{align}
where $e$ and $a$ are eccentricity and semi-major axis of the ellipse, respectively, with
\begin{align}
&\label{eq:eccentricityCentral}
e = \sqrt{\beta} ~ , \\
&\label{eq:semimajorCentral} a = \frac{K}{E}\qty[\qty(\frac{p_\theta c}{K})^2 - 1] \frac{1}{1 - e^2} ~ , \\
&\kappa \equiv \qty[1 - \qty(\frac{K}{p_\theta c})^2]^{1/2} ~ , \\
& \beta \equiv \qty(\frac{m_0 c^2}{E})^2 + \qty(\frac{p_\theta c}{K})^2 - \qty(\frac{m_0 c^2}{E}\frac{p_\theta c}{K})^2 ~ .
\end{align}

\begin{figure}
	\begin{center}
		\includegraphics[width=6.8cm]{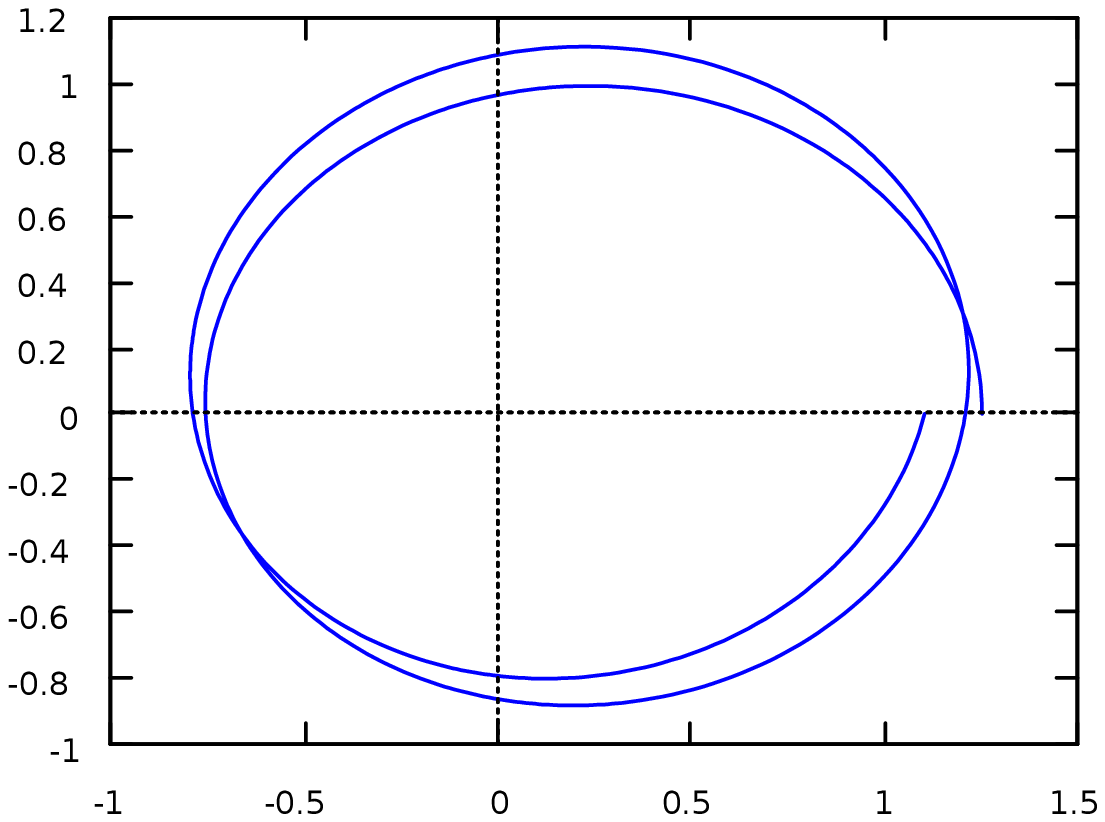}
	\end{center}
	\caption{\label{fig:graph1}Case $e = 1/4$, $a = 1$}
	\begin{center}
		\includegraphics[width=6.8cm]{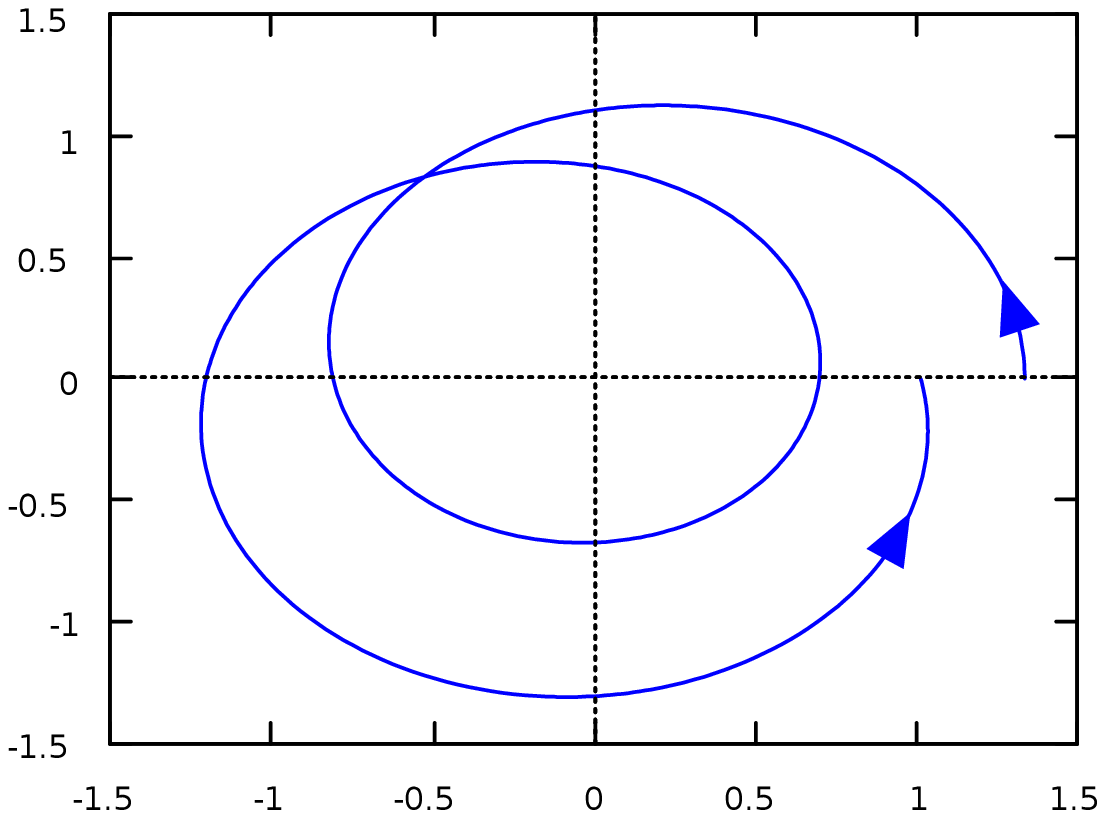}
	\end{center}
	\caption{\label{fig:graph2}Case $e = 1/3$, $a = 1$}
	\begin{center}
		\includegraphics[width=6.8cm]{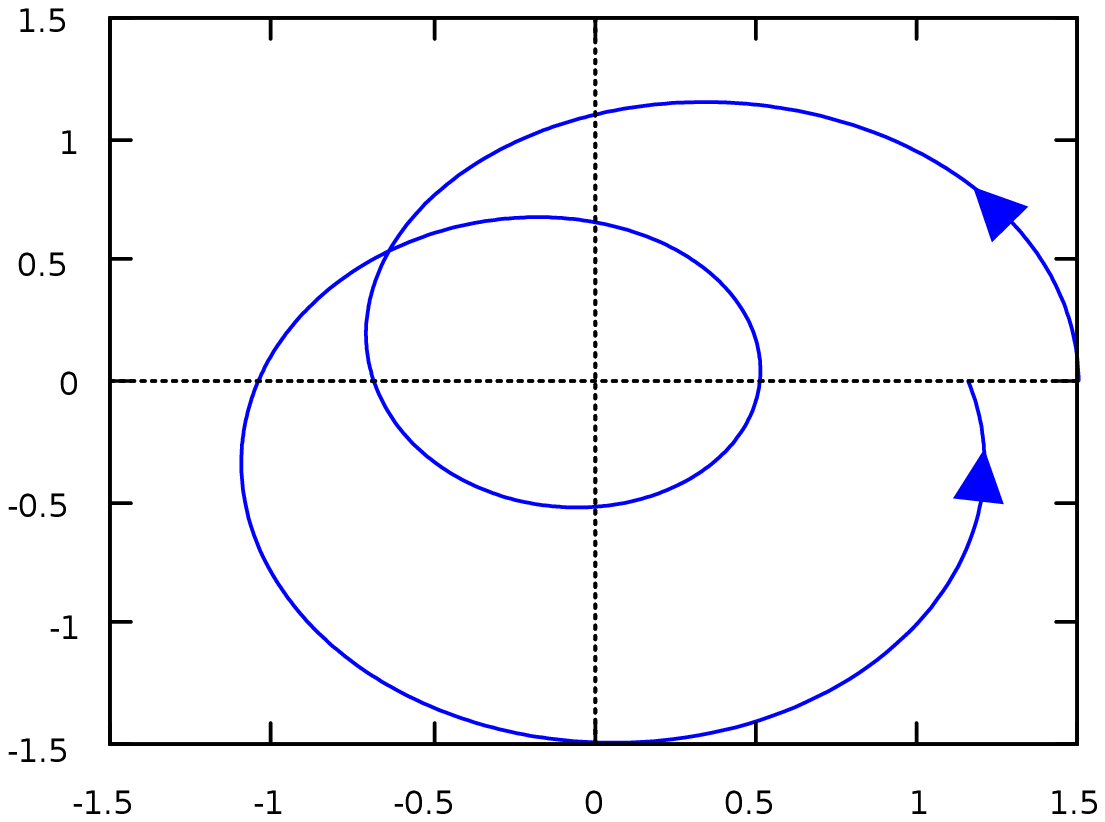}
	\end{center}
	\caption{\label{fig:graph3}Case $e = 1/2$, $a = 1$}
\end{figure}

We could express the angular momentum and the energy in terms of $a$ and $e$,  namely,
\begin{align}
&\label{eq:angularMomentumSolved}\frac{p_\theta^2 c^2}{K^2} =  \frac{1+e^2}{2}\qty[1 + \frac{1 - e^2}{1 + e^2}\sqrt{1 + 4\eta^2}]  ~ ,  \\
& E = \frac{K}{a(1 - e^2)}\qty[\frac{p^2_\theta  c^2}{K^2} - 1] ~ ,
\end{align}
respectively, with $\eta \equiv \frac{m_0 c^2 a}{K}$. It is also important to state that in this relativistic case we have the orbital precession
\begin{align}
2\pi\delta = \frac{2\pi}{1 - \frac{K^2}{2 p_\theta^2 c^2}} - 2\pi ~ .
\end{align}
As a numerical simulation, we set $m = K = 1$. For $e = 1/4, a = 1$, the orbit of mass is shown in the figures \ref{fig:graph1}, \ref{fig:graph2}, and \ref{fig:graph3}.
In the case of $K/p_\theta c \ll 1$, we could have $\kappa \approx 1 - \frac{K^2}{2 p_\theta^2 c^2}$ so that
\begin{align}
\delta = \qty(\frac{2p_\theta^2 c^2}{K^2} - 1)^{-1} \approx \frac{K^2}{2 p_\theta^2 c^2} ~ .
\end{align}

We would like to write some comments as follows. Although the orbital precession can be regarded as relativistic effect, it is of less interest to consider it at the level of microscopic physics since it is almost impossible to observe this effect. Instead, we have relativistic quantum mechanics for this level. Therefore, it is more interesting to discuss it at the level of macroscopic physics, for example, involving a case of  gravitational interaction in the special theory of relativity.

\section{ Hydrogen-like Atomic Model}
\label{sec:HydrogenAtom}

\begin{figure}
	\begin{center}
		\includegraphics[width=10cm]{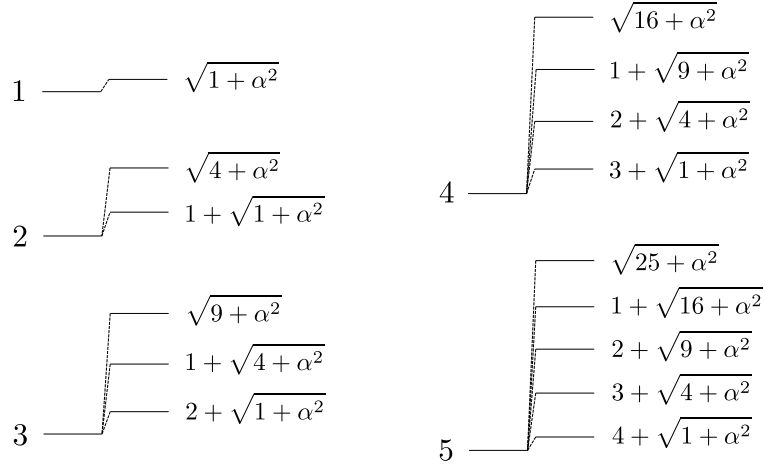}
	\end{center}
	\caption{\label{fig:levelDiagram}The values of $n'$. Each branch of clusters value is approximately equal to the parent branch.}
\end{figure}

In this section we apply our study in the preceding case to the model of hydrogen-like atom where an electron with mass $m_e$ orbits a proton as nucleus with mass $m_N$. They interact by the Coulomb force with $K = q^2/4\pi\epsilon_0$, where $q$ is the magnitude of electron and proton charges. This case at hand can be viewed as a single particle model moving under the influence of the Coulomb force with reduced mass  $\mu = \qty(1/m_e + 1/m_N)^{-1}$.  Using Wilson-Sommerfeld rule of quantization
\begin{align}
& \oint p_\theta\dd \theta = n_\theta h, \rightarrow p_\theta\int_0^{2\pi}\dd\theta = n_\theta h ~ , \\
& \oint p_r\dd r = n_r h \rightarrow p_\theta \oint \frac{u'^2}{u^2}\dd \theta = p_\theta \oint \frac{\kappa^2 \sin[2](\kappa\theta)}{\qty(1/e + \cos(\kappa\theta))^2}\dd\theta  = n_r h  ~,
\end{align}
with $n_\theta, n_r \in \mathbb{Z}$. From the first integral, we have
\begin{align}
p_\theta = n_\theta \hbar ~ ,
\end{align}
whereas for the second integral, it gives
\begin{align}
\label{eq:eccenQuantization}
\int_0^{2\pi} \frac{ \sin[2](\tilde\theta)}{\qty(1/e + \cos(\tilde\theta))^2}\dd\tilde\theta = \frac{2\pi}{\sqrt{1-e^2}} + 2\pi = \frac{2\pi n_r}{\sqrt{n_\theta^2 - \alpha^2}} ~ ,
\end{align}
 where $\tilde\theta = \kappa\theta$, $\kappa = \sqrt{1 - (\alpha^2/n_\theta^2)}$, and $\alpha = q^2/(4\pi\epsilon_0\hbar c)$ implying
\begin{align}
e = \qty[1 - \qty(\frac{\sqrt{n_\theta^2 - \alpha^2}}{n_r + \sqrt{n_\theta^2 - \alpha^2}})^2]^{1/2}.
\end{align}
Since $e \in [0,1]$, $n_\theta$ cannot be zero since it produces complex number for $e$, then $n_\theta \geq 1$. While $n_r$ can be chosen $n_r \geq 0$. Defining $\tilde{n} \equiv \sqrt{n_\theta^2 - \alpha^2}$ and $n' \equiv n_r + \sqrt{n_\theta^2 - \alpha^2}$, we have
\begin{align}
e = \sqrt{1 - \frac{\tilde{n}^2}{n'^2}} ~ .
\end{align}
The values of $n'$ is given in  Figure \ref{fig:levelDiagram}.

From equation \eqref{eq:eccentricityCentral}, the eccentricity is given by
\begin{align}
& e = \pm \qty[\qty(\frac{\mu c^2}{E})^2 + \qty(\frac{n_\theta}{\alpha})^2 - \qty(\frac{n_\theta}{\alpha}\frac{\mu c^2}{E})^2]^{1/2} ~ .
\end{align}
The energy of the system is
\begin{align}
E &=  \mu c^2\sqrt{\frac{n_\theta^2 - \alpha^2}{n_\theta^2 - e^2\alpha^2}}
=  \frac{\mu c^2}{\sqrt{1 + \dfrac{\alpha^2}{n'^2}}}
=  \mu c^2\qty(1 + \dfrac{\alpha^2}{\qty(n_r + \sqrt{n_\theta^2 - \alpha^2})^2})^{-1/2}.
\end{align}
The above energy formula resembles the Sommerfeld's fine structure \cite{Granovskii2004SommerfeldFA, Lu1970DerivationOS, Terzis2008ASR} (see also,\cite{greiner_relativistic_2000} p. 231)
\begin{align}
&E = mc^2\qty(1 + \frac{\alpha^2}{\qty[n - j - \frac{1}{2} + \sqrt{(j+\frac{1}{2})^2 - \alpha^2}]^2})^{-1/2}, \notag \\
& n = 1,2,3,\dots, \quad j = \frac{1}{2},\frac{3}{2},\frac{5}{2},\dots
\end{align}
where $n$ is the principal quantum number, and $j$ is the total angular momentum number. We conclude  $j = n_\theta - \frac{1}{2}, n = n_r + j + \frac{1}{2} = n_r + n_\theta$.

\section{The Orbital Precession in Modified Gravity}
\label{sec:OrbitPrecession}

This section is devoted to consider a solar system where a planet with mass $m$ moves around a massive star with mass $M$. We assume that the orbit is closed. In this section, we take $K = GMm$ where $G$ is the universal gravitational constant.

First, let us focus on a modified gravitational theory whose potential energy has the form
\begin{align}
\label{eq:PotentialPerturbed}
V(r) = -\frac{K}{r} + \frac{\epsilon K_1}{r^3} ~ ,
\end{align}
where $K_1$ is a real constant and $\epsilon $ is a small dimensionless quantity. The above form of potential energy is motivated by generalized Binet equation derived using Schwarzschild geometry, see for example \cite{cheng2009} and  a modified gravitational theory in the context of Randall-Sundrum Model \cite{Randall_1999}.  It is important to notice that since $K_1$ is an arbitrary real constant  in \eqref{eq:PotentialPerturbed}, it also covers the case in \cite{Lemmon2010} where they have fixed the constant coming from perturbative modified gravitational force.

Setting $\epsilon \ll 1$, the Binet equation \eqref{eq:MainBinet2} simplifies to
\begin{align}
\label{eq:BinetForPrecession}
u'' + \kappa^2 u - \frac{E K}{p_\theta^2 c^2} = -\epsilon K_1 \qty(\frac{4 K}{p_\theta^2 c^2}u^3 + \frac{3 E }{p_\theta^2 c^2}u^2) + \mathcal O(\epsilon^2) ~ ,
\end{align}
such that we can write the solution of \eqref{eq:BinetForPrecession}  as
\begin{align}
u(\theta) = u_0(\theta) + \epsilon u_1(\theta) ~ . \label{eq:ansatzsol}
\end{align}
The function $u_0$ is the unperturbed solution given in \eqref{eq:solutionBinetCentral}, whereas the function $u_1$ is the first order of perturbation with the boundary value  $u_1(0)=u_1'(0)=0$. Inserting the ansatz \eqref{eq:ansatzsol} into \eqref{eq:BinetForPrecession}, it results
\begin{align}
u_0'' + \epsilon u_1'' + \kappa^2(u_0 + \epsilon u_1) - \frac{E K}{p_\theta^2 c^2} = -\epsilon K_1\qty(\frac{4 K}{p_\theta^2 c^2}u_0^3 + \frac{3 E}{p_\theta^2 c^2}u_0^2) + \mathcal O(\epsilon^2) ~ .
\end{align}
Then, equalizing the first order terms, we have
\begin{align}
u_1'' + \kappa^2 u_1 = -u_0^2\qty(\frac{4 KK_1}{p_\theta^2 c^2}u_0 + \frac{3 EK_1}{p_\theta^2 c^2}) ~ ,
\end{align}
whose solution has the form
\begin{align}
u_1 = c_1\cos(\kappa\theta) + c_2\sin(\kappa\theta) + v_1(\theta)\cos(\kappa\theta) + v_2(\theta)\sin(\kappa\theta) ~ ,
\end{align}
where $c_1,c_2$ are real constants, and
\begin{align}
\label{eq:v1}
v_1(\theta) &= \frac{K_1}{\kappa p_\theta^2 c^2}\int_0^\theta\dd \theta'\ u_0^2\qty(4Ku_0 + 3E)\sin(\kappa\theta') \notag\\
&= -\frac{K_1}{ea\kappa^2 p_\theta^2 c^2\qty(1 - e^2)^2}\bigg[
\frac{K\qty[1 + e\cos(\kappa\theta)]^4}{ea\qty(1 - e^2)^3} + E\qty[1 + e\cos(\kappa\theta)]^3
\bigg] ~ ,\\
\label{eq:v2}v_2(\theta) &= -\frac{K_1}{\kappa p_\theta^2 c^2}\int_0^\theta\dd \theta'\ u_0^2\qty(4Ku_0 + 3E)\cos(\kappa\theta') \notag \\
& = -\frac{K_1}{8a\kappa^2 p_\theta^2 c^2\qty(1 - e^2)^3} \bigg[
Ke^3\sin(4\kappa\theta) + \qty{12Eea(1-e^2) + 8Ke^3 + 24 Ke}\sin(2\kappa\theta) \notag \\
&\quad + \qty{-8Ee^2a(1-e^2) - 32Ke^2}\sin[3](\kappa\theta)
+ \qty{(24Ee^2 + 24E)a(1-e^2) + 96Ke^2 + 32K}\sin(\kappa\theta) \notag \\
&\quad + \qty{24Eea(1-e^2) + 12Ke^3 + 48 Ke}\kappa\theta
\bigg],
\end{align}
where $e$ and $a$ are given in \eqref{eq:eccentricityCentral} and \eqref{eq:semimajorCentral} respectively. According to the initial value, $u_1(0) = u_1'(0) = 0$, we should have $c_1 = -v_1(0)$ and $c_2 = -v_2(0) = 0$. Therefore, the complete orbit function is
\begin{align}
\label{eq:OrbitPrecession}
u(\theta) = \frac{1 + e\cos(\kappa\theta)}{a(1 - e^2)} - v_1(0)\epsilon\cos(\kappa\theta) + \epsilon v_1(\theta)\cos(\kappa\theta) + \epsilon v_2(\theta)\sin(\kappa\theta) ~ .
\end{align}

From the equations \eqref{eq:OrbitPrecession}, \eqref{eq:v1}, and \eqref{eq:v2}, it is easy to see that $v_1$ is bounded from above and below since it consists of cosinus functions, thus we can ignore the term $\epsilon v_1\cos(\kappa\theta)$. We can also ignore the term $v_1(\theta) \epsilon \cos(\kappa\theta)$ by the same reason. Meanwhile,  $v_2$,  is a non bounded function on $\theta$, so that we can approximate the last term in  \eqref{eq:OrbitPrecession} to
$$-\frac{\epsilon K_1 \qty{24Eea(1-e^2) + 12Ke^3 + 48 Ke}\theta\sin(\kappa\theta)}{8a\kappa p_\theta^2 c^2(1 - e^2)^3} ~ .$$
Thus, we have
\begin{align}\label{eq:usolaprox}
u(\theta) \approx \frac{1 + e\cos(\kappa\theta)}{a(1 - e^2)}  -\frac{\epsilon K_1}{\kappa p_\theta^2 c^2}\frac{6Eea(1-e^2) + 3Ke^3 + 12 Ke}{2a(1 - e^2)^3}\theta\sin(\kappa\theta) ~ .
\end{align}
Let us define
\begin{align}
\sigma \equiv -\frac{\epsilon K_1}{\kappa p_\theta^2 c^2}\frac{6Eea(1-e^2) + 3Ke^3 + 12 Ke}{2a(1 - e^2)^3} ~ .
\end{align}
Since $\sigma$ is small, then
\begin{align}
\alpha\cos(\kappa\theta) + \sigma\theta \sin(\kappa\theta)
\approx  \alpha\cos(\qty(\kappa - \frac{\sigma}{\alpha})\theta) ~,
\end{align}
where
\begin{align}
\alpha = \frac{e}{a(1 - e^2)} ~ .
\end{align}
So we can simplify \eqref{eq:usolaprox} to
\begin{align}
u(\theta) = \frac{1 + e\cos(\qty[\kappa - \frac{\sigma a(1 - e^2)}{e}]\theta)}{a(1 - e^2)} ~ .
\end{align}
The orbital precession when the body revolves in one period
\begin{align}
2\pi\delta = \frac{2\pi}{\kappa - \frac{\sigma a(1 - e^2)}{e}} - 2\pi ~ .
\end{align}
Taking the approximation
\begin{align}
 \frac{K^2}{ p_\theta^2 c^2} \ll 1 ~ ,
\end{align}
we finally have the orbital precession
\begin{align}\label{eq:deltagen}
\delta = \frac{K^2}{2p_\theta^2 c^2}  \qty[1 - \frac{3 \epsilon K_1}{K(1-e^2)^2} \qty(  \frac{2  p_\theta^2 c^2 }{K^2} + e^2 + 2)]  ~ .
\end{align}

\subsection{Comparison with  Other Theories}
We consider an example of Mercury's orbit. In this case $K = GMm$, where $G$ is Newton gravity constant, $M = 1.989\times 10^{30}\ \mathrm{kg}$ is mass of Sun, $m = 3.301\times 10^{23}\ \mathrm{kg}$ is mass of Mercury. Its semi-major axis $a = 57.91\times 10^9 \mathrm{m}$ and eccentricity $e=0.2056$. Since
\begin{align}
\frac{m c^2}{GMm} \approx 10^{-3} > \frac{e}{a(1-e^2)}  \approx 10^{-8}  ~ ,
\end{align}
and  $\eta \gg e$, then eq. \eqref{eq:angularMomentumSolved} becomes approximately
\begin{align}\label{eq:GML}
& \frac{p_\theta^2 c^2}{(GMm)^2} = \frac{ c^2 a(1 - e^2)}{GM} ~ .
\end{align}
Inserting \eqref{eq:GML} into \eqref{eq:deltagen}, we obtain
\begin{align}
\delta &= \frac{GM}{2c^2 a(1 - e^2)} - \frac{3\epsilon K_1}{2 m c^2a(1-e^2)^3 }\qty( \frac{2 c^2 a(1-e^2)}{GM} + e^2 +2) ~ .
\end{align}
Meanwhile, in general relativity the orbit precession of Mercury is given by \cite{cheng2009}
\begin{align}
\delta_{GR} = \frac{3 GM}{c^2a(1 - e^2)} = \epsilon  \approx 10^{-7}  .
\end{align}
In order to make a contact with observation, we have to set $\delta = \delta_{GR}$ implying
\begin{align}
 K_1 = -\frac{5a (1-e^2)^3 GMmc^2}{9 \qty(GM(2 + e^2) + 2 a c^2 (1-e^2)) } ~ .
\end{align}

Another interesting gravitational theory is Randall-Sundrum model \cite{Randall_1999}. In this theory  we have a five dimensional bulk spacetime with negative cosmological constant in which we embed two 3-branes (3+1 dimensional spacetimes) lying on the fifth coordinate with distance $ r_c$ where $r_c$ is called compactification radius. After integrating out all massive Kaluza-Klein particles by taking $ r_c$ to be very large (compared to Planck length $\sim 10^{-35}$ m), we finally obtain the modified Newtonian gravitational potential energy
\begin{align}
\label{eq:PotentialRS}
V(r) = -\frac{K}{r} - \frac{\epsilon_{RS} K}{r^3} ~ ,
\end{align}
where $\epsilon_{RS} \propto 1/k^2$ is a dimensionless quantity and $k$ is a warped factor on the bulk spacetime. After employing similar computation as above, we obtain the orbit precession of Mercury
\begin{align}\label{eq:deltaRS}
\delta_{RS} &= \frac{GM}{2c^2 a(1 - e^2)} + \frac{3 \epsilon_{RS} G M}{2  c^2a(1-e^2)^3 }\qty( \frac{2 c^2 a(1-e^2)}{GM} + e^2 +2) ~ .
\end{align}
However, since the warped factor $k$ is propotional to Planck energy in SI unit, that is, $k \approx 10^8$, so the last term in \eqref{eq:deltaRS} is suppressed to zero. Thus, this Randall-Sundrum model is ruled out in this case.

At the end, it is worth mentioning that in Newtonian theory we could also use the potential energy \eqref{eq:PotentialPerturbed}. Using the same reason as above, we have to take
\begin{align}
 K_1 = - \frac{a^2}{3} (1 - e^2)^2 GMm ~ ,
\end{align}
in order to have consistent model with observation.

\subsection{Relevance with second Kepler's law}
It is well known that in the problem of planetary orbit, the motion should satisfy the second Kepler's law. Let us consider a section area in the orbit plane $\dd A = \frac{1}{2}r^2\dd \theta$. Then, it can be shown that the rate of swept area is constant since the $\gamma$ factor and the angular momentum $ p_\theta$ are constant.  In other words, $\dot A$ is constant, namely,
\begin{align}
\label{eq:rateOfArea}
\dot A = \frac{p_\theta}{2\gamma m_0} = \frac{a(1-e)}{2m_0\sqrt{m_0^2 c^2 + p_\theta^2 a^2(1-e)^2}} ~ .
\end{align}

\begin{figure}
	\begin{center}
		\includegraphics[width=6.8cm]{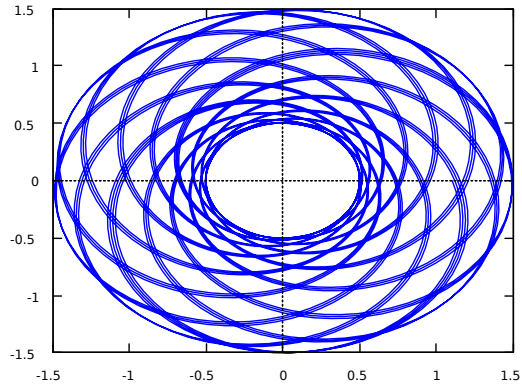}
	\end{center}
	\caption{\label{fig:graph4}The orbit of a planet shifted by precession.}
\end{figure}

The area of ellipse is $\pi \sqrt{1-e^2} a^2$. From \eqref{eq:rateOfArea}, the Kepler's law in this case has the form
\begin{align}
\int_0^T\dd t\ \dot A = \frac{p_\theta T}{2\gamma m_0} = n\pi \sqrt{1-e^2} a^2 ~ ,
\end{align}
where $T$ is the period such that the orbit of a planet  is closed and $n$ is the number revolution of a planet forming closed orbit which has to be a positive integer as illustrated by the Figure \ref{fig:graph4}.

\section{Conclusion}
We have discussed the motion of a body under the influence of central force in the special theory of relativity. We particularly reviewed the motion under the force that is proportional to $1/r^2$ where $r$ is the radial coordinate in which it admits the precessing elliptical orbit. This orbital precession depends on the eccentricity and semi-major axis of the orbit. If the orbit is circular, then the orbital precession does not appear.

Then, we applied the above results to the case of the hydrogen-like atom. We used the Wilson-Sommerfeld quantization rule to obtain the energy level of the system so called Sommerfeld-Dirac fine structure formula which can also be derived from Dirac equation in relativistic quantum mechanics.

At the end, we have considered the planetary motion in which the potential energy is fine-tuned by adding the term which is proportional to $1/r^3$. We obtained the explicit orbital function and then, derived the precession of the planetary orbit. Since the pre-coefficient of the additional term in the potential energy is arbitrary, we could make a contact with the result from general relativity. However, in the Randall-Sundrum model this pre-coefficient is suppressed to zero implying that the model can be ruled out. Also, we pointed out the second Kepler's laws which differs to the case of Newtonian mechanics.

\ack

We are grateful to Fiki Taufik Akbar,  Ardian Nata Atmaja, Andy Octavian Latief, and Rudy Kusdiantara  for useful discussion and valuable suggestions. The work in this paper is supported by World Class Professor program of Kemendikbud Ristek 2021.

\appendix
\section{\label{app:ConstantOfMotion} Derivation of Constant of Motion} 
For a relativistic particle motion under the influence of  central potential, the Lagrangian is given by
\begin{align}
L = -\frac{m_0 c^2}{\gamma(v)} - V(r), \quad \gamma = \qty(1 - \frac{\dot r^2}{c^2} - \frac{r^2 \dot \theta^2}{c^2})^{-1/2}.
\end{align}
The conjugate momenta correspond to $r,\theta$ are 
\begin{align}
& p_r = \gamma m_0 \dot r ~ , \\
& p_\theta = \gamma m_0 r^2\dot\theta ~ ,
\end{align}
respectively. From Euler-Lagrange equation of motions for $\theta$, we have
\begin{align}
\dv{t}\qty(\gamma m_0 r^2 \dot \theta) = 0 ~ ,
\end{align}
where the quantity $\gamma m_0 r^2 \dot\theta$ is the angular momentum $p_\theta$. From the above equation, it implies that $p_\theta$ is constant.

Now we consider the total time derivative of Lagrangian,
\begin{align}
\dot L = \pdv{L}{r} \dot r + \pdv{L}{\theta} \dot \theta + \pdv{L}{\dot r}\ddot r + \pdv{L}{\dot\theta}\ddot\theta + \pdv{L}{t}.
\end{align}
The last term vanish since the Lagrangian does not depend explicitly on time. Meanwhile,
\begin{align}
\pdv{L}{\dot q_i}\ddot q_i = \dv{t}\qty(\pdv{L}{\dot q_i}\dot q_i) - \dv{t}\qty(\pdv{L}{\dot q_i})\dot q_i,\quad \text{where} \quad q_i = r,\theta ~ .
\end{align}
Then,
\begin{align}
\dot L = \qty[\pdv{L}{r} - \dv{t}\qty(\pdv{L}{\dot r})]\dot r + \qty[\pdv{L}{\theta} - \dv{t}\qty(\pdv{L}{\dot \theta})]\dot \theta + \dv{t}\qty(\pdv{L}{\dot r}\dot r) + \dv{t}\qty(\pdv{L}{\dot \theta}\dot \theta) ~ . 
\end{align}
From Euler-Lagrange equation of motions, the first and second terms vanish implying
\begin{align}
\dv{t}\qty(\pdv{L}{\dot r}\dot r + \pdv{L}{\dot\theta}\dot\theta - L) = 0 ~ ,
\end{align}
which is nothing but the energy function of the system. Thus,  we conclude that the total energy of the system
\begin{align}
E = \pdv{L}{\dot r}\dot r + \pdv{L}{\dot\theta}\dot\theta - L = \gamma(v) m_0 c^2 + V(r) ~ ,
\end{align}
is also the constant of motion.

\section{\label{app:DerivationBinetInverseSquare} Orbit Equation}
We begin by writing the Binet equation 
\begin{align}
\qty(\dv[2]{\theta} + 1 - \frac{K^2}{p_\theta^2 c^2})u = \frac{KE}{p_\theta^2 c^2} ~ .
\end{align}
The characteristic equation of the above differential equation for homogeneous case is
\begin{align}
\rho^2 + 1 - \frac{K^2}{p_\theta^2 c^2} = 0 ~ ,
\end{align}
whose solution is given by
\begin{align}
\rho = \pm i\Omega, \quad \Omega = \sqrt{1-\frac{K^2}{p_\theta^2 c^2}} ~ .
\end{align}
Let $u_0$ be the solution to the homogeneous equation
\begin{align}
u_0 = A\exp(i\Omega \theta) + B\exp(-i\Omega\theta) ~,
\end{align}
whereas $u_1$ is the particular solution. Since the rhs of the Binet equation is constant, then  $u_1'' = 0$ such that we have

\begin{align}
u_1 = \frac{KE}{p_\theta^2 c^2 - K^2} ~ .
\end{align}
Finally, we can write the solution as
\begin{align}
u(\theta) = u_0(\theta) + u_1 = A\exp(i\Omega \theta) + B\exp(-i\Omega\theta) + \frac{KE}{p_\theta^2 c^2 - K^2} ~ .
\end{align}

\section{Derivation of Eccentricity}
Let us first consider
\begin{align}
\qty(m_0 c^2)^2 = \qty(E + Ku)^2 - u'^2p_\theta^2 c^2 - u^2p_\theta^2 c^2 ~ ,
\end{align}
for $\theta = 0$ which implies
\begin{align}
\qty(m_0 c^2 a(1-e))^2 = \qty(Ea(1-e) + K)^2 - \qty(p_\theta c)^2 ~ .
\end{align}
Defining $\alpha = a(1-e)$, we get
\begin{align}
\qty(m_0^2 c^4 - E^2)\alpha^2 - 2 K E\alpha - \qty(K^2 - p_\theta^2 c^2) = 0 ~ ,
\end{align}
resulting
\begin{align}
\alpha = \frac{KE}{m_0^2 c^4 - E^2}\qty[1 \pm \sqrt{\qty(\frac{m_0 c^2}{E})^2 + \qty(\frac{p_\theta c}{K})^2 - \qty(\frac{m_0 c^2}{E}\frac{p_\theta c}{K})^2}] ~ .
\end{align}
Since
\begin{align}
a = \frac{K}{E}\qty(\frac{p_\theta^2 c^2}{K^2} - 1)\frac{1}{1 - e^2} ~ , \label{eq:sbmayor}
\end{align}
it gives then
\begin{align}
\frac{K}{E}\qty(\frac{p_\theta^2 c^2}{K^2} - 1)\frac{1}{1 + e} = \frac{KE}{m_0^2 c^4 - E^2}\qty[1 \pm \sqrt{\qty(\frac{m_0 c^2}{E})^2 + \qty(\frac{p_\theta c}{K})^2 - \qty(\frac{m_0 c^2}{E}\frac{p_\theta c}{K})^2}] ~,
\end{align}
whose solution is 
\begin{align}
e =  \sqrt{\beta} ~,
\end{align}
where
\begin{align}
\beta = \qty(\frac{m_0 c^2}{E})^2 + \qty(\frac{p_\theta c}{K})^2 - \qty(\frac{m_0 c^2}{E}\frac{p_\theta c}{K})^2 ~ . \label{eq:beta}
\end{align}

\section{Solving the angular momentum}

Let $x = p_\theta c/K$, then we substitute \eqref{eq:sbmayor} into \eqref{eq:beta} we have
\begin{align}
(x^2 - 1)\qty[(x^2 - e^2)(x^2 - 1) - \eta^2(1 - e^2)^2] = 0 ~ ,
\end{align}
whose solutions are given by
\begin{align}
x^2 \in \qty{1, \frac{1+e^2}{2}\qty[1 + \frac{1 - e^2}{1 + e^2}\sqrt{1 + 4\eta^2}], \frac{1+e^2}{2}\qty[1 - \frac{1 - e^2}{1 + e^2}\sqrt{1 + 4\eta^2}]} ~ , 
\end{align}
where $\eta \equiv \frac{m_0 c^2 a}{K}$. Note that in the third solution,  $\eta \geq \frac{e}{1 - e^2}$ must hold.

\section{Solving The Radial Momentum Quantization}
We start from
\begin{align}
& n \int_0^{2\pi} \frac{u'^2}{u^2}\dd \theta = 2\pi k ~ ,
\end{align}
such that
\begin{align}
\int_0^{2\pi} \qty(n\frac{u'^2}{u^2} - k) \dd \theta = 0 ~.
\end{align}
Since $u'^2/u^2 = \kappa^2 e \sin[2](\kappa\theta)/(1 + e\cos(\kappa\theta))^2$, $\kappa^2 = 1 - (\alpha^2/n^2)$, then
\begin{align}
\frac{\qty(n - (\alpha^2/n)) e \sin[2](\kappa\theta)}{(1 + e\cos(\kappa\theta))^2} = k ~ .
\end{align}
We use the above equation to solve the relation between $e$, $k$, and $n$. For $\theta = \pi/2$, we have 
\begin{align}
\qty(n - \frac{\alpha^2}{n}) e = k ~ ,
\end{align}
and for $\theta = \pi/4$, we have
\begin{align}
\frac{\qty(n - \frac{\alpha^2}{n}) e}{2(1 + \qty(e/\sqrt{2}))^2} = k ~.
\end{align}

\end{document}